\def\ii#1\ff{\textul{#1}}	

\def\bbsty#1#2#3{{\bf #1}, (#3) #2}	

\documentclass[psfig,amsmath,amssymb]{cimento_modif}
\usepackage{graphicx}\usepackage{epsfig}
\usepackage{amsmath}\usepackage{amssymb}\usepackage{amsfonts}
\usepackage{bm}
\usepackage{exscale}
\usepackage{dcolumn}    
\usepackage{color}
\usepackage{soul}
\usepackage{ulem}
\shorttitle{Interplay between surface and volume instabilities in heavy-ion collisions}
\title{Interplay between surface and volume instabilities in heavy-ion collisions examined within mean-field extensions}
\author{
	P.~Napolitani\from{ins:x},	
	H.~Dinh Viet\from{ins:x}\ETC,	\atque
	M.~Colonna\from{ins:y}	}
\instlist{
	\inst{ins:x} Universit\'e Paris-Saclay, CNRS/IN2P3, IJCLab, 91405 Orsay, France
	\inst{ins:y} INFN-LNS, Laboratori Nazionali del Sud, 95123 Catania, Italy}
 
\begin{document}

\maketitle

%
%
\begin{abstract}
	In the transition from nuclear matter to finite nuclei, complex finite-size effects which characterise open systems arise, in relation with either the nuclear surface or the bulk. In addition, the non-equilibrium character of the process, typical of violent heavy-ion collisions (from Fermi energy to the intermediate-energy domain) adds up as well. 
	The resulting dynamics is the combination of surface and volume unstable modes which trigger large-amplitude fluctuations. 
	A rich variety of fragmentation patterns may emerge, ranging from collimated streams of nuclear clusters to the split of a stretched nuclear complex into few large fragments. 
	They imply different conditions of density and surface tension, and result in different chronologies. 
	Such phenomenology has been observed in experiments, but it is often difficult to recognise and disentangle the underlying types of instabilities.
	To draw some example, two extremely deformed nuclear systems, produced below and above Fermi energy, are chosen and followed microscopically all along their evolution within the Boltzmann-Langevin One-Body approach.
\end{abstract}

%
%
%
%
%
\section{Introduction: two extremely deformed situations}
%


	As far as nuclear-matter conditions are matched, like in some astrophysical situations, the evolution of a nuclear system can be efficiently described in terms of equation of state.
	In this case, the disintegration of the system into fragments and clusters is associated to extreme thermal conditions or to Coulomb-frustration phenomena, as in the case of dense stellar matter.
	Macroscopically, this corresponds to the nuclear liquid-gas phase-transitions and, more specifically, to a situation of mechanical instability (or negative incompressibility), where fluctuations can be amplified in the bulk.

	This simple thermodynamic view, which is matched in several astrophysical scenarios, holds in the limited conditions of very large systems and in circumstances where the timescale of the process largely exceeds the relaxation time of the nuclear interaction.
	However, the understanding of the nuclear process may turn into a complex puzzle when moving from nuclear matter conditions to finite open systems, as in the case of heavy-ion collisions.
	In this situation, first of all, the initial stages of heavy-ion collisions may progress in so short timescales that chaotic behaviours dominate.
	Secondly, a surface is introduced and the corresponding surface instabilities come into play.
	With respect to nuclear-matter conditions, where volume instabilities characterise the bulk, the contribution of the surface and of a non-trivial geometry is a major modification.
	These conditions call for a dynamic description, where no equilibrium assumptions are required and where volume and surface fluctuations combine in determining a rich variety of evolution paths, large-amplitude fluctuations and, ultimately, the splitting of the system into nuclear fragments and clusters.

	In this respect, we analyse two opposite scenarios within a common microscopic picture and focus on the associated dynamic observables.
	These are, firstly, extremely deformed unstable heavy systems formed in deep-inelastic collisions and, secondly, collimated streams of clusters, called nuclear jets, produced in mass-asymmetric head-on collisions at Fermi energies.
	Comparing these two cases is particularly instructive because, without an accurate analysis, they would be source of misinterpretation and even present apparent, though misleading, similarities.
	We rely on the Boltzmann-Langevin One Body (BLOB) model~\cite{Napolitani2013,Napolitani2017},
where the interplay of collective and dissipative modes is explicitly handled.

\section{From deep inelastic to nuclear jets beyond Fermi energy with BLOB}

	In dissipative heavy-ion collisions, the observables which could be studied experimentally vary as a function of the incident energy, especially around Fermi energy.
	The two above-mentioned case studies are situated below and above Fermi energy.

\begin{figure}[t!]\begin{center}
	\includegraphics[angle=0, width=1\textwidth]{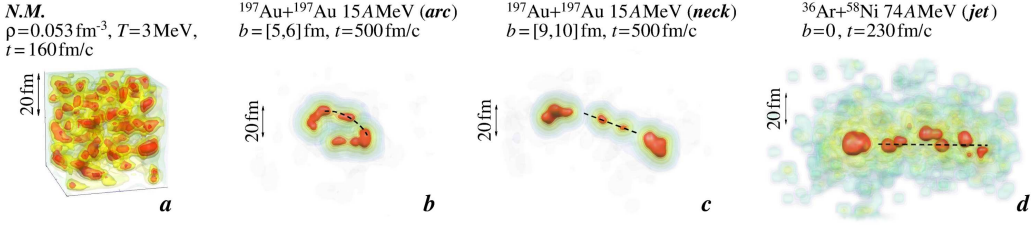}
\end{center}\caption
{
	Large-amplitude fluctuations in nuclear matter and in three finite nuclear systems (BLOB calculations).
	{\bf a}, Nuclear matter in unstable condition showing mottling~\cite{Napolitani2017}.
	{\bf b}, $^{197}$Au$+^{197}$Au at 15~$A\textrm{MeV}$ (semi-peripheral), fragmentation of an arc thread~\cite{Napolitani2019a}.
	{\bf c}, $^{197}$Au$+^{197}$Au at 15~$A\textrm{MeV}$ (peripheral), fragmentation of a neck~\cite{Napolitani2019a}.
	{\bf d}, $^{36}$Ar$+^{58}$Ni at 74~$A$MeV (head-on), nuclear jet~\cite{Napolitani2019b}.
}
\label{fig1}
\end{figure}

	Below Fermi energy, deep-inelastic regimes are generally associated with collective modes.
	We may take as an example the system $^{197}$Au$+^{197}$Au at 15~$A\textrm{MeV}$, which has been measured~\cite{SkwiraChalot2008,Wilczynski2010b,Wilczynski2010a}.
	In this case, extremely deformed and unstable composite systems can be produced when nuclei which are too heavy to undergo fission are involved.
	Such heavy systems, progressing from a very deformed configuration, have been found to go through fast splits into three or even four fragments which are not mere isotropic emissions around the target or projectile remnants.
	This mechanism has been observed even for non-peripheral collisions.
	Two types of descriptions are usually proposed.
	Either, instabilities act on two separate stages, so that residues result from fast secondary fissions of well separated and still deformed projectile-like and target-like fragments.
	Or, instabilities act on the whole system at once, so that residues emerge promptly in the midrapidity region between the projectile-like and target-like fragments.
	The system $^{197}$Au$+^{197}$Au, inspired a few modelling attempts, within molecular dynamics (QMD simulation~\cite{Tian2010}) or Boltzmann approaches (Stochastic mean-field simulation~\cite{Rizzo2014}), where some approximations concern either the description of mean field or fluctuations, respectively.
	While chronology and impact-parameter dependencies are equally handled in both simulations, clear differences stem out when fluctuations and collective effects are combined.
	Within molecular dynamics, deformations and collective effects are inhibited according to an approximated mean field; as a result, the configurations produced in the collision are too compact and fractal-like.
	Within a Boltzmann approach, even with the inclusion of fluctuations in an approximated form (as the case of the SMF model), fragmentation is inhibited by mean-field resilience; as a result, the configurations produced in the collision are very elongated and too hard to break.
	On the other hand, a description based on the BLOB approach succeeds in getting closer to the experimental results~\cite{Napolitani2019a}.
	Fig.~\ref{fig1}a illustrates the mottling configuration of a periodic portion of nuclear matter calculated with BLOB in mechanically unstable condition, while fig.~\ref{fig1}b and fig.~\ref{fig1}c show two configurations achieved in $^{197}$Au$+^{197}$Au collisions where arc and neck threads form and break into fragments.
	In a first preliminary statement, we may suggest that the BLOB approach joins the advantages of the previous approaches in providing an accurate description of the mean field and in describing fluctuations in the full phase space.
	A more accurate analysis, addressed below, will focus on the instability conditions.
%
%

	At Fermi energy and above, two-body dissipation dominates and is associated with the emission of several clusters and fragments.
	We take as an example head-on $^{36}$Ar$+^{58}$Ni collisions which have been measured~\cite{Lautesse2006,Francalanza2017} at various bombarding energies in a range from 25 to 95~$A\textrm{MeV}$.
	Such mechanism, where head-on collisions take place between nuclei of asymmetric size, leads to the disintegration of the lighter partner into a nuclear jet, i.e. a collimated stream of clusters, while the heavier partner cools down in a more gentle evaporation process.
	In this case, the description of a nuclear jet requires to model a rapid out-of-equilibrium clusterisation process,
where both surface tension and two-body collisions compete.
	In other words, it is fundamental to handle an accurate description of the mean field while handling the fluctuation-dissipation process in full phase space, which is quite the same requirement as for the first case study. 
	A description based on the BLOB approach has in fact proven to be successful in reproducing the mechanism as it was observed experimentally, an illustration is presented in fig.~\ref{fig1}d.
	Variations from the BLOB approach, like the suppression of the two-body collision contribution, would lead to different results which where not observed, like the development of a neck between a projectile-like and a target-like fragment.
	On the other hand, early statistical interpretations (not suited to out-of-equilibrium conditions) attributed the process to vaporisation~\cite{Bacri1995,Borderie1996}, supposing that very large temperatures could be reached.
	If the process actually tends to produce several light charged particles at larger bombarding energies, given its highly out-of-equilibrium character, a more explanatory analysis should rather be focused on identifying the type of instability that rules this process, and it requires a linear-response analysis. 

	These two scenarios, at least due to the energy regime they are associated to, deep-inelastic and Fermi energies, are expected to be associated to substantially different underlying instabilities, even though fragments are produced in both cases.
	Both of them could be described efficiently within a common approach relying on the BLOB model, but they might be difficult to characterise on the basis of usual production and kinematic observables.
	In this spirit, we further exploit the common microscopic simulation framework in order to access observables related to the process.
	In particular, we focus on a linear-response analysis, which is able to describe the type and the properties of the involved instabilities in terms of the associated grow rate and fluctuation amplitude.


\section{Instabilities and related observables}

%
%
\begin{figure}[b!]\begin{center}
	\includegraphics[angle=0, width=1\textwidth]{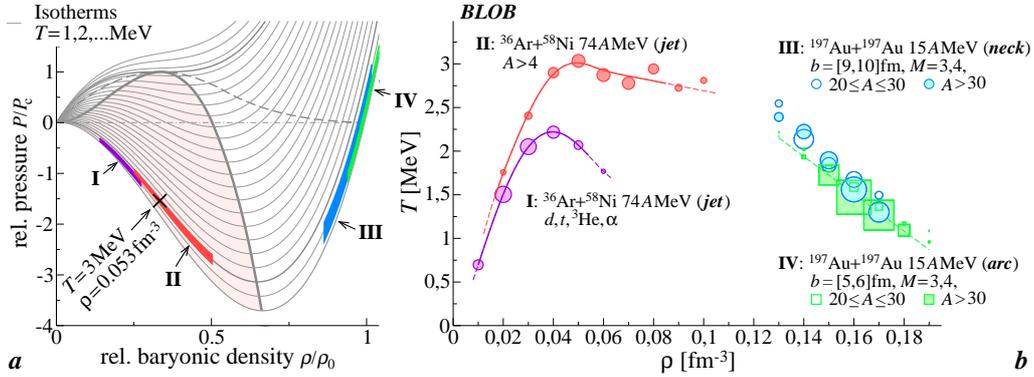}
\end{center}\caption
{	Thermal and density conditions associated to the following mechanisms, calculated with BLOB.
Production of $d$, $t$, $^3$He and $\alpha$ (I) or fragments with $A>4$ (II) in a nuclear jet in $^{36}$Ar$+^{58}$Ni head-on collisions at 74~$A$MeV.
	Fragments from the rupture of arc or neck threads in $^{197}$Au$+^{197}$Au at 15~$A\textrm{MeV}$ in semi-peripheral (IV) and peripheral (III) collisions.
	{\bf a}, temperature and density explored in I,II,III and IV indicated on the nuclear equation-of-state landscape.
	{\bf b}, temperature and density explored by cluster and fragments at the moment of their formation in I,II,III. Symbol sizes are proportional to the total yields of the reaction channels (I and II combined, III or IV)
}
\label{fig2}
\end{figure}

	As advocated, a microscopic analysis of the instabilities which brings the system to break into fragments and clusters calls for retracing the transition from nuclear matter to the conditions of a finite self-bound open system.
	In nuclear matter, the nuclear interaction leads to clusterisation below saturation density and in unstable sites of the equation of state, called spinodal~\cite{Chomaz2004}, typically corresponding to a negative incompressibility (i.e. density increasing with decreasing pressure, which impries mechanical instability).
	The possibility of extracting thermal and density observables from the microscopic model as a function of time allows to compare the two systems of our case studies on the same basis and to construct the corresponding dispersion relations for both volume and surface instabilities.
	Fig.~\ref{fig2} tracks temperature-versus-density correlations of forming fragments. 
	These quantities are extracted from the kinetic-energy distribution and the density landscape in the formation site and at the time when fragments start to emerge.
	In deep inelastic $^{197}$Au$+^{197}$Au semi-peripheral and peripheral collisions at 15~$A\textrm{MeV}$, fragments are formed along neck or arc threads, respectively.
	The density oscillates slightly around saturation, so that conditions are insufficient for processes depending on density gradients and, in general, rupture conditions for fragment formation cannot be associated to volume instabilities.

	Completely different clusterisation conditions characterises the nuclear jets produced in $^{36}$Ar$+^{58}$Ni at 74~$A\textrm{MeV}$.
	The density reaches about $1/3$  $\rho_{\textrm{sat}}$ for intermediate-mass fragments with $A>4$ and drops even below, at about $1/5$  $\rho_{\textrm{sat}}$ for $d$, $t$, $^3$He and $\alpha$ clusters (see ref.~\cite{Napolitani2019b} for details).
	The temperature, extracted when fragments are emerging and when the process should have approached some condition of equilibrium, is in the range that usually characterises multifragmentation processes.
	In this respect, the thermal properties of this mechanism, even when light clusters are produced, correspond entirely to the spinodal region of the equation-of-state landscape and do not seem to match vaporisation conditions.

\begin{figure}[b!]\begin{center}
	\includegraphics[angle=0, width=1\textwidth]{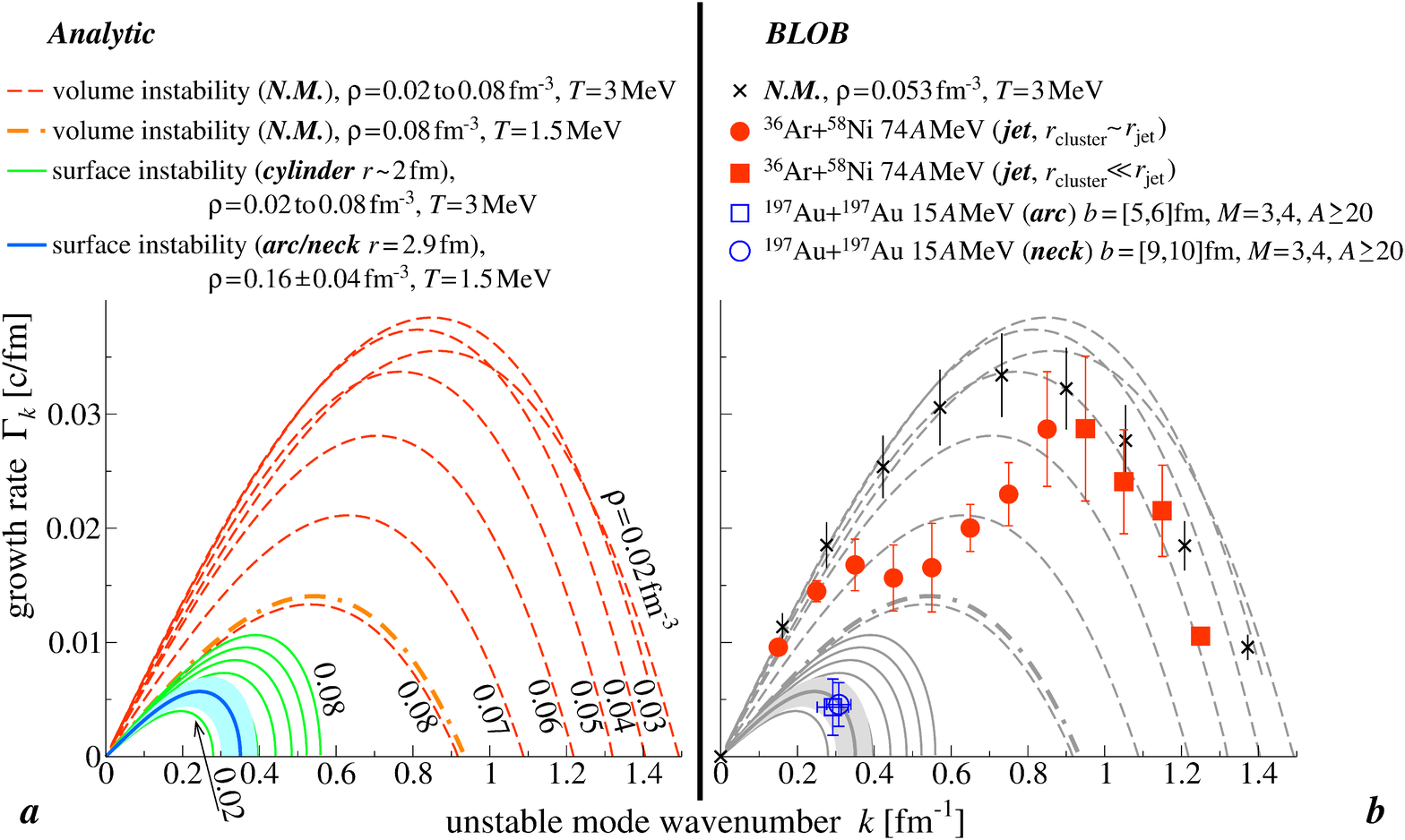}
\end{center}\caption
{
	Dispersion relations to compare volume and surface instabilities for nuclear matter, along neck and arc threads in the systems $^{197}$Au$+^{197}$Au at 15~$A\textrm{MeV}$ in semi-peripheral and peripheral collisions and in nuclear jets produced in $^{36}$Ar$+^{58}$Ni head-on collisions at 74~$A$MeV.
	{\bf a}, Analytic calculations of volume instabilities (spinodal) in nuclear matter and analytic study of surface (Plateau-Rayleigh) instabilities along threads which mimic the topologies obtained in the selected finite systems.
	{\bf b}, Numerical calculations for nuclear matter conditions and the selected finite systems with BLOB.
}
\label{fig3}
\end{figure}

	The spinodal conditions, as illustrated in fig.~\ref{fig3}, can be described within a dispersion relation displaying the growth rate $\Gamma_k$ at which the fluctuation-response amplitude of a volume-unstable mode is amplified as a function of the corresponding wave number $k$ (see ref.~\cite{Napolitani2017} for details).
	For small $k$ modes (or large wavelengths $\lambda=1/k$) the growth rate $\Gamma_k$ is small (or the growth time $\tau_k=1/\Gamma_k$ is large).
	Also for large $k$ modes, or small wavelengths, the growth rate $\Gamma_k$ is small and drops to zero as the
nuclear interaction imposes an ultraviolet cutoff, excluding small wavelengths $\lambda$.
	The resulting one-hump-shaped dispersion relation presents a maximum for the leading $k$ modes which manifest the largest growth rate and define the most probable spinodal wavelength in the unstable process. 

	When moving from nuclear matter 
to open systems, the leading $k$ modes correspond to the most probable spinodal fragment size.
	At $\rho_{\textrm{sat}}/3$ in nuclear matter the leading $k$ modes correspond to a spinodal wavelength of about $8\,$fm; in an open system the corresponding spinodal fragment size is typically in the range from O to Ne.
	In the evolution of the process in nuclear matter, possible interferences may appear at later times so that larger $k$ modes recombine into smaller $k$ modes; correspondingly, in an open system, heavier elements arise from the recombination of smaller inhomogeneities in the density landscape, as triggered by mean-field resilience.
	Differently from nuclear matter, open systems introduce also a surface, with the related deformations and instabilities, so that surface instabilities~\cite{Brosa1990} (of Plateau-Rayleigh type) should be added to the volume (spinodal) instabilities.
	In terms of the same interaction which already determined the dispersion relation for the spinodal modes, we can also define a dispersion relation for the Plateau-Rayleigh modes (see ref.~\cite{Napolitani2019a} for details).
	This latter describes the growth rate of a surface instability which characterises a columnar thread, which may correspond to a neck topology produced in peripheral heavy-ion collisions or in the early configuration of a very stretched system that is going to break into a stream of collimated nuclear clusters (i.e. a nuclear jet).
	
%
%

	In order to quantify the role of instabilities and the contribution of cohesive properties we should carry out a large-amplitude response analysis.
	Such analysis is shown in fig.~\ref{fig3}, where dispersion relations for volume and surface modes are compared for the two case studies.
	For the system $^{197}$Au$+^{197}$Au at 15~$A\textrm{MeV}$, the microscopic BLOB simulation tracks the fluctuation growth rate $\Gamma_k$ as a function of the corresponding wavenumber $k$ along arcs and neck threads at the average temperature obtained for the rupture conditions (studied in fig.~\ref{fig2}). 
	In particular, the growth rate is obtained as $\Gamma_k=\hbar/t_{\textrm{rupture}}$, where $t_{\textrm{rupture}}$ is the average time when the threads break into fragments, counted with respect to the instant when inhomogeneities start to arise.
	In order to attribute this result to given types of instabilities, we compare to the corresponding analytic descriptions.
	Since for this system the density presents only a small variation around saturation, spinodal conditions are simply not matched in the density-temperature equation of state landscape for the average temperature obtained for the rupture conditions.
	Still, for indicative purposes, we can calculate the spinodal dispersion relation in nuclear matter for $\rho_{\textrm{sat}}/2$, and we confirm that it does not match with the numeric result: the leading wavelength $\lambda\sim20$~fm, which is associated to very massive fragments (up to Ge) is twice larger than the leading spinodal wavelength, while the growth time, reaching values of about 200 to 300 fm$/$c is up to ten times larger than the growth time of the spinodal leading mode in nuclear matter.
	We calculate analytically the dispersion relation for the Plateau-Rayleigh modes at various densities at rupture temperature and we confirm that it matches the numeric result at saturation density.
	For this system we can therefore conclude that any fragment configuration, either aligned or annular-like, arises from the rupture of a very deformed thread around saturation density due to surface instabilities of Plateau-Rayleigh type.

	For the system $^{36}$Ar$+^{58}$Ni at 74~$A\textrm{MeV}$ we follow the same procedure as for the previous system, addressed this time to the forward side of nuclear jet, where clusters are formed.
	The numerical result can be compared to analytic prescriptions for the spinodal dispersion relation in nuclear matter and for the Plateau-Rayleigh dispersion relation along a columnar thread corresponding to the nuclear jet at rupture temperature.
	We find a completely different result with respect to the previous deep-inelastic situation.
	Even though clusters emerge from a very elongates thread, the dispersion relation is situated in the spinodal region and far outside the Plateau-Rayleigh region, suggesting that clusters are mainly produced from volume instabilities.
	A combination of different density modes may result in favouring smaller wavelengths and, as a consequence, in enhancing the production of lighter clusters.
	For smaller wavenumbers, a contribution from surface fluctuation may also add up producing a reduction of the growth rate $\Gamma_k$ in the dispersion relation.

\section{Conclusions}

	Quite generally, the initial stages of heavy-ion collisions progress in so short timescales that chaotic behaviours dominate.
	Explanatory theories for finite systems should in this case not progress from equilibrium assumptions but rather allow for a microscopic analysis of unstable conditions and the associated modes, which characterise the surface and volume of the system.
	We tested such approach on two selected systems, $^{197}$Au$+^{197}$Au at 15~$A\textrm{MeV}$, which is dominated by surface instabilities and $^{36}$Ar$+^{58}$Ni at 74~$A\textrm{MeV}$, which is mostly dominated by volume instabilities.

	The first system, $^{197}$Au$+^{197}$Au at 15~$A\textrm{MeV}$, despite matching the deep-inelastic regime, undergoes multiple splits, even in misaligned configurations. 
	We can conclude from the microscopic analysis that these exotic mechanisms are actually produced by extreme deformations (necks, arcs and annular threads) associated with surface instabilities around saturation density, and are not compatible with an expansion processes.
	They are also not compatible with mechanism observed at Fermi energy where nuclear necks are associated to low densities and to density-driven isospin drifts.

	In the second system, $^{36}$Ar$+^{58}$Ni at 74~$A\textrm{MeV}$, head-on collisions produce a nuclear jet which disintegrates into a swarm of clusters.
	The jet is however not compatible with a neck thread resulting from cohesive forces (like in peripheral collisions at Fermi energy).
	Low densities are explored and surface tension becomes negligible, so that spinodal-like volume instabilities actually dominate, even though we have shown that some residual effect from surface instabilities should also be taken into account in the dispersion relation.
	The result is a rapid clusterisation into collimated streams of light nuclei and mostly $\alpha$ particles which is however fully compatible with spinodal instability, without reaching the high temperatures which characterise vaporisation.


\acknowledgments
Research was conducted under the auspices of the International Research Program (IRP) COLL-AGAIN.


\begin{thebibliography}{0}
%
%
%
%
%
%

%
\bibitem{Napolitani2013} 
\BY{Napolitani P. \atque Colonna M.}
\textit{Phys. Lett. B} \bbsty{726}{382}{2013}.
%
\bibitem{Napolitani2017} 
\BY{Napolitani P. \atque Colonna M.}
\textit{Phys. Rev. C} \bbsty{96}{054609}{2017}.
%
\bibitem{SkwiraChalot2008}
\BY{Skwira-Chalot I. et al.} 
\textit{Phys. Rev. Lett.} \bbsty{101}{262701}{2008}.   
%
\bibitem{Wilczynski2010b}
\BY{Wilczy\'nski J. et al.}
\textit{Phys. Rev. C} \bbsty{81}{024605}{2010}.   
%
\bibitem{Wilczynski2010a}
\BY{Wilczy\'nski J. et al.}
\textit{Phys. Rev. C} \bbsty{81}{067604}{2010}.
%
\bibitem{Tian2010}
\BY{Junlong Tian et al.}
\textit{Phys. Rev. C} \bbsty{82}{054608}{2010}.   
%
\bibitem{Rizzo2014}
\BY{Rizzo C., Colonna M., Baran V. \atque Di Toro M.}
\textit{Phys. Rev. C} \bbsty{90}{054618}{2014}.
%
\bibitem{Napolitani2019a}
\BY{Napolitani P., Sainte-Marie A. \atque  Colonna M.}
\textit{Phys. Rev. C} \bbsty{100}{054614}{2019}.
%
\bibitem{Napolitani2019b}
\BY{Napolitani P. \atque  Colonna M.}
\textit{Phys. Lett. B} \bbsty{797}{134833}{2019}.
%
\bibitem{Lautesse2006}
\BY{Lautesse P. et al. (INDRA coll.)}
\textit{Eur. Phys. J.} \bbsty{A27}{349}{2006}.
%
\bibitem{Francalanza2017} 
\BY{Francalanza L. et al.}
\textit{IOP Conf. Ser.} \bbsty{863}{012061}{2017}.
%
\bibitem{Bacri1995} 
\BY{Bacri Ch.O. et al.}
\textit{Phys. Lett. B} \bbsty{353}{27}{1995}.
%
\bibitem{Borderie1996} 
\BY{Borderie B. et al.}
\textit{Phys. Lett. B} \bbsty{388}{224}{1996}.
%
\bibitem{Chomaz2004}
\BY{Chomaz Ph., Colonna M. \atque Randrup J.}
\textit{Phys. Rep.} \bbsty{389}{263}{2004}.
%
\bibitem{Brosa1990}
\BY{Brosa U., Grossmann S. \atque M\"uller A.}
\textit{Phys. Rep.} \bbsty{197}{167}{1990}.




\end{thebibliography}
\end{document}